\RequirePackage{lineno}
\documentclass[aps,prl,twocolumn,noeprint,nofootinbib]{revtex4-1}
\usepackage{xcolor}
\usepackage{graphics}
\usepackage{graphicx}
\usepackage[normalem]{ulem}
\usepackage{soul,xcolor}
\newcommand{\agev}    {\mbox{$A$~GeV}}               
\newcommand{\gevc}    {\mbox{GeV$/c$}}

\newcommand{\rb}[1]   {\mbox{\textrm{\scriptsize #1}}}
\newcommand{\rbt}[1]  {\mbox{\textrm{\tiny #1}}}
\newcommand{\hefour}  {\ensuremath{{}^{4}\textrm{He}}}

\newcommand{\sqrtsnn} {\ensuremath{\sqrt{s_{_{\rbt{NN}}}}}}

\newcommand{\pt}      {\ensuremath{p_{\rb{t}}}}

\newcommand{\dedx}    {\ensuremath{\textrm{d}E/\textrm{d}x}}

\newcommand{\ebeam}   {\ensuremath{E_{\rb{beam}}}}
\newcommand{\yproj}   {\ensuremath{y_{\rb{proj}}}}
\newcommand{\ycm}     {\ensuremath{y_{\rb{cm}}}}
\newcommand{\etacm}   {\ensuremath{\eta_{\rb{cm}}}}

\newcommand{\vn}      {\ensuremath{v_{n}}}
\newcommand{\vone}    {\ensuremath{v_{1}}}
\newcommand{\vtwo}    {\ensuremath{v_{2}}}
\newcommand{\vthree}  {\ensuremath{v_{3}}}
\newcommand{\vfour}   {\ensuremath{v_{4}}}
\newcommand{\vfive}   {\ensuremath{v_{5}}}
\newcommand{\vsix}    {\ensuremath{v_{6}}}

\newcommand{\psirp}   {\ensuremath{\Psi_{\rb{RP}}}}
\newcommand{\psiep}   {\ensuremath{\Psi_{\rb{EP}}}}
\newcommand{\psieone} {\ensuremath{\Psi_{\rb{EP,1}}}}
\begin{document}
%
%

\title{Directed, elliptic and higher order flow harmonics
  of protons,
  deuterons and tritons in Au+Au collisions at $\sqrtsnn = 2.4$~GeV}

%
\author{
  J.~Adamczewski-Musch$^{5}$,
  O.~Arnold$^{11,10}$,
  C.~Behnke$^{9}$,
  A.~Belounnas$^{17}$,
  A.~Belyaev$^{8}$,
  J.C.~Berger-Chen$^{11,10}$,
  A.~Blanco$^{2}$,
  C.~Blume$^{9}$,
  M.~B\"{o}hmer$^{11}$,
  P.~Bordalo$^{2}$
% ...
}
\affiliation{
  (HADES collaboration) \\
  \\
  \mbox{$^{1}$Institute of Physics, Slovak Academy of Sciences,
    84228~Bratislava, Slovakia} \\
  \mbox{$^{2}$LIP-Laborat\'{o}rio de Instrumenta\c{c}\~{a}o e
    F\'{\i}sica Experimental de Part\'{\i}culas,} \\
  \mbox{3004-516~Coimbra, Portugal} \\
  \mbox{$^{3}$Institute of Nuclear Physics, Polish Academy of
    Sciences, 31342 Krak\'{o}w, Poland} \\
  \mbox{$^{4}$Smoluchowski Institute of Physics, Jagiellonian University
    of Cracow,} \\
  \mbox{30-059~Krak\'{o}w, Poland} \\
  \mbox{$^{5}$GSI Helmholtzzentrum f\"{u}r Schwerionenforschung GmbH,
    64291~Darmstadt, Germany} \\
  \mbox{$^{6}$Technische Universit\"{a}t Darmstadt, 64289~Darmstadt,
    Germany} \\
  \mbox{$^{7}$Institut f\"{u}r Strahlenphysik, Helmholtz-Zentrum
    Dresden-Rossendorf, 01314~Dresden, Germany} \\
  \mbox{$^{8}$Joint Institute of Nuclear Research, 141980~Dubna,
    Russia} \\
  \mbox{$^{9}$Institut f\"{u}r Kernphysik, Goethe-Universit\"{a}t,
    60438 ~Frankfurt, Germany} \\
  \mbox{$^{10}$Excellence Cluster 'Origin and Structure of the
    Universe', 85748~Garching, Germany} \\
  \mbox{$^{11}$Physik Department E62, Technische Universit\"{a}t
    M\"{u}nchen, 85748~Garching, Germany} \\
  \mbox{$^{12}$II.Physikalisches Institut, Justus Liebig
    Universit\"{a}t Giessen, 35392~Giessen, Germany} \\
  \mbox{$^{13}$Institute for Nuclear Research, Russian Academy of
    Science, 117312~Moscow, Russia} \\
  \mbox{$^{14}$Institute of Theoretical and Experimental Physics,
    117218~Moscow, Russia} \\
  \mbox{$^{15}$National Research Nuclear University MEPhI (Moscow
    Engineering Physics Institute),} \\
  \mbox{115409~Moscow, Russia} \\
  \mbox{$^{16}$Department of Physics, University of Cyprus,
    1678~Nicosia, Cyprus} \\
  \mbox{$^{17}$Laboratoire de Physique des 2 infinis Ir\`{e}ne
    Joliot-Curie, Universit\'{e} Paris-Saclay,} \\
  \mbox{ CNRS-IN2P3., F-91405 Orsay, France} \\
  \mbox{$^{18}$Nuclear Physics Institute, The Czech Academy of
    Sciences, 25068~Rez, Czech Republic} \\
  \mbox{$^{19}$LabCAF. F. F\'{\i}sica, Univ. de Santiago de
    Compostela, 15706~Santiago de Compostela, Spain} \\
  \mbox{$^{20}$Uniwersytet Warszawski, Wydzia\l\ Fizyki, Instytut
    Fizyki Do\'{s}wiadczalnej, 02-093~Warszawa, Poland} \\
  \mbox{$^{21}$Dipartimento di Fisica and INFN, Universit\`{a}
    di Torino, 10125~Torino, Italy} 
  \\
  \\
  \mbox{$^{a}$ also at Coimbra Polytechnic - ISEC, ~Coimbra, Portugal}
  \\
  \mbox{$^{b}$ also at Technische Universit\"{a}t Dresden,
    01062~Dresden, Germany} \\
  \mbox{$^{c}$ also at Frederick University, 1036~Nicosia, Cyprus} \\
  \mbox{$^{\dagger}$ deceased} \\
} 
%
%
\date{\today}
\begin{abstract}
Flow coefficients \vn\ of the orders $n = 1 - 6$ are measured with the
High-Acceptance DiElectron Spectrometer (HADES) at GSI for protons,
deuterons and tritons as a function of centrality, transverse momentum
and rapidity in Au+Au collisions at $\sqrtsnn = 2.4$~GeV.  Combining
the information from the flow coefficients of all orders allows to
construct for the first time, at collision energies of a few GeV, a
multi-differential picture of the angular emission pattern of these
particles.  It reflects the complicated interplay between the effect
of the central fireball pressure on the emission of particles and
their subsequent interaction with spectator matter.  The high
precision information on higher order flow coefficients is a major
step forward in constraining the equation-of-state of dense baryonic
matter.
\end{abstract}
%
%
\maketitle
%

%
Heavy-ion collisions in the center-of-mass energy range of $\sqrtsnn
\approx 1 - 10$~GeV provide access to the properties of strongly
interacting matter at very high net-baryon densities, which also
define the characteristics of astrophysical objects like neutron
stars \cite{Adamczewski-Musch:2019byl}.  Important information on this
form of matter, e.g. on its Equation-Of-State (EOS), can be inferred
from the measurement of collective flow 
\cite{Danielewicz:2002pu,Fevre:2015fza}.  The majority of the flow
studies at SIS18 and AGS energies performed up to now were restricted
to the analysis of directed and elliptic flow (for a review see
\cite{Ritter:2014uca,Andronic:2006ra,Herrmann:1999wu,Reisdorf:1997fx}).
These correspond to the first (\vone) and second (\vtwo) order
coefficients of the Fourier decomposition \cite{Voloshin:1994mz} of
the azimuthal angle $\phi$ distribution of emitted particles with
respect to the orientation of the reaction plane (RP).  The latter is
defined by the beam axis $\vec{z}$ and the direction of the impact
parameter $\vec{b}$ of the colliding nuclei, which is given by the RP
angle \psirp\footnote{In the following, $\pt = \sqrt{p_{x}^{2} +
    p_{y}^{2}}$ is the transverse momentum and $y = 1 / 2 \, \ln [(E +
  p_{z}) / (E - p_{z}) ]$ the rapidity of a given particle in the
  laboratory frame.  The rapidity in the center-of-mass system is
  denoted by $\ycm = y - 1 / 2 \, \yproj$, with the projectile
  rapidity $\yproj = 1.48$.}.
It has been shown that important information can be extracted from an
analysis of higher order flow coefficients relative to \psirp.  For
instance, a comparison of the proton \vthree\ measured by HADES with
UrQMD transport model calculations indicates that in particular
\vthree\ exhibits an enhanced sensitivity to the EOS of the hadronic
medium \cite{Hillmann:2018nmd,Hillmann:2019wlt}.  Other transport
model calculations suggest that a non-vanishing fourth order
coefficient (\vfour) measured at center-of-mass energies of a few GeV
can constrain the nuclear mean field at high net-baryon densities
\cite{Danielewicz:1999zn}.  At high energies (RHIC and LHC) the
measurements of higher order flow coefficients relative to the
symmetry plane of identical order were decisive to determine the shear
viscosity over entropy density $\eta/s$ of QCD matter at high
temperatures \cite{Heinz:2013th}.  Attempts have also been made to
extract $\eta/s$ for dense hadronic matter at lower energies by
employing transport models
\cite{Demir:2008tr,Khvorostukhin:2010aj,Barker:2016hqv,Rose:2017bjz}
or hydrodynamic approaches \cite{Ivanov:2016hes}.  Since these studies
did not converge on conclusive results yet, input from measurements of
higher order flow coefficients at low energies will be essential to
further constrain the theoretical descriptions.  Important information
can be derived from an analysis of the scaling properties of higher
flow harmonics.  Initial theoretical considerations suggested, e.g., a
simple scaling of \vtwo\ and \vfour\ as $\vfour(\pt) / \vtwo^{2}(\pt)
= 1/2$ for an ideal fluid scenario \cite{Borghini:2005kd}, while later
measurements at RHIC \cite{Adams:2003zg,Adare:2010ux} and LHC
\cite{ATLAS:2012at,CMS:2013bza,Acharya:2018lmh} have revealed a more
complex behavior.  In the few GeV center-of-mass energy range, the
flow pattern is strongly affected by the presence of slow spectator
nucleons.  They interfere with the particle emission from the central
fireball and cause a distinct evolution of the relative contribution
of odd and even flow harmonics as a function of rapidity
\cite{Reisdorf:1997fx,Ritter:2014uca}.

In this letter we report first measurements of higher order flow
harmonics (i.e. $v_{n}$ with $n = 3, 4, 5$ and $6$) for protons,
deuterons and tritons in fixed-target Au+Au collisions at $\ebeam =
1.23$\agev, corresponding to a center-of-mass energy in the
nucleon-nucleon system of $\sqrtsnn = 2.4$~GeV.
%

%
The HADES experiment consists of six identical detection sections
located between the coils of a toroidal superconducting magnet which
each cover polar angles between $18^{\circ}$ and $85^{\circ}$,
corresponding to the center-of-mass pseudo-rapidity range $-0.79 <
\etacm < 0.96$, and almost $\pi/3$ in azimuth.  Each sector is
equipped with a Ring-Imaging Cherenkov (RICH) detector followed by
four layers of Multi-Wire Drift Chambers (MDCs), two in front of and
two behind the magnetic field, as well as a Time-Of-Flight detector
(TOF) ($44^{\circ}$~--~$85^{\circ}$) and Resistive Plate Chambers
(RPC) ($18^{\circ}$~--~$45^{\circ}$).  Hadrons are identified using
the time-of-flight measured with TOF and RPC and the energy-loss
information from TOF, as well as from the MDCs.  Their momenta are
determined via the deflection of the tracks in the magnetic field.
The event plane (EP) angle is calculated from the emission angles and
charges of projectile spectators as measured in the Forward Wall (FW)
detector.  It consists of 288 scintillator modules which are read out
by photomultiplier tubes.  The FW is placed at a $6.8$~m distance from
the target and covers the polar angles $0.34^{\circ} < \theta <
7.4^{\circ}$.  The minimum bias trigger is defined by a signal in a
60~$\mu$m thick mono-crystalline diamond detector (START)
\cite{Pietraszko:2014tba}, which is positioned in the beam line.  In
addition, online Physics Triggers (PT) are used based on hardware
thresholds on the TOF signal corresponding to at least 5 (PT2) or 20
(PT3) hits in the TOF detector.  By comparing the measured TOF+RPC hit
multiplicity distribution with Glauber Model simulations it has been
estimated that the PT3~trigger is selecting about $43$~\%
(PT2~trigger: $72$~\%) of the total inelastic cross section of $6.83
\pm 0.43$~barn \cite{Adamczewski-Musch:2017sdk}.  This multiplicity is
also used for the offline centrality determination.  For this analysis
the PT3~triggered event sample is divided into four centrality
intervals, each corresponding to $10$~\% of the total Au+Au cross
section.  A detailed description of the HADES experiment can be found
in Ref.~\cite{Agakishiev:2009am}.  

Tracks are reconstructed using the hit information of the MDCs and
Particle IDentification (PID) is based on their time-of-flight.
Protons, deuterons and tritons are selected within windows of $2.5
\cdot \sigma_{\beta}(p)$ width around the corresponding particle
velocity $\beta$ expected for a given momentum $p$.  The resolutions
$\sigma_{\beta}(p)$ also depend on $p$ and are parameterized
accordingly.  To suppress contaminations to the particle sample
identified via time-of-flight, in particular the \hefour~contribution
to the deuteron sample, the energy loss (\dedx) measurements in the
MDCs are used in addition.  Phase space regions with a PID purity
below 80~\% are excluded from the analysis.  In high multiplicity
Au+Au collisions reconstruction efficiencies depend on the local track
multiplicities.  Since collective effects will cause anisotropies of
the event shape, corresponding to local variations of the track
densities and thus of the reconstruction efficiencies, a data-driven
correction procedure depending on the track orientation relative to
the EP is applied.

\begin{figure}
\includegraphics[width=1.\linewidth]{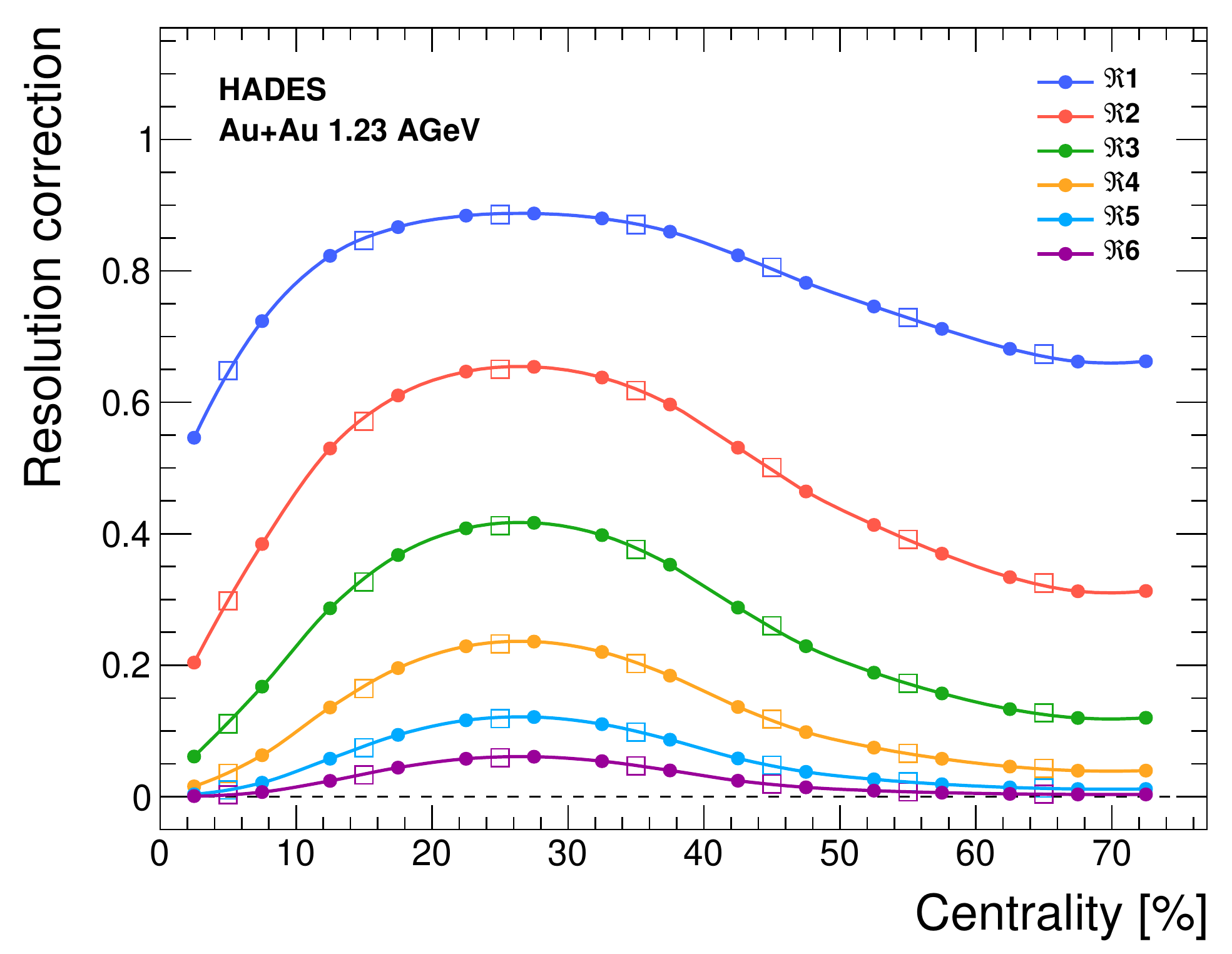}
\caption{The resolution of the first order spectator event plane
  $\Re_{n}$ for the flow harmonics of different orders $n$ as a
  function of the event centrality.  The circles correspond to
  centrality intervals of $5$~\% width and the squares to $10$~\%
  width (curves are meant to guide the eye).}
\label{fig:EPresolution}
\end{figure}

In the analysis presented here the azimuthal distributions of particle
yields relative to the azimuthal orientation of the RP is used to
determine the flow coefficients \vn\
\cite{Ollitrault:1993ba,Ollitrault:1997vz,Poskanzer:1998yz}.
However, as the azimuthal angle of the RP \psirp\ is not accessible to
measurements, an estimator for this angle, the EP angle \psiep\ is
introduced.  For its determination hits of projectile spectators in
the FW are used.  From the laboratory angles $\phi_{FW}$ of the 
fired FW cells a vector $\vec{Q}_{n} = (Q_{n,x}, Q_{n,y}) = (\sum w \;
\cos (n \, \phi_{FW}) , \sum w \; \sin (n \,\phi_{FW}) \,)$ of order
$n$ is calculated event-by-event.  As weights the charges $w = |Z|$
are used, as determined from the signal height measured in a given FW
cell.  Non-uniformities in the FW acceptance and a possible
misalignment of the beam are corrected by applying the standard
re-centering method \cite{Poskanzer:1998yz} to the positions
$X_{\rb{FW}}$ and $Y_{\rb{FW}}$ by shifting the first moments
$(\langle X_{\rb{FW}} \rangle, \langle Y_{\rb{FW}} \rangle)$ and
dividing them by the second moments
$(\sigma_{X_{\rb{FW}}},\sigma_{Y_{\rb{FW}}})$.  Residual
non-uniformities in the EP angular distribution are removed
by an additional flattening procedure \cite{Barrette:1997pt}.  The
first order EP angle is then given by $\psieone = \arctan(Q_{1, y} /
Q_{1, x} )\, $.  The flow coefficients of all orders discussed here
are defined relative to \psieone, i.e. the first order EP measured via
the spectator nucleons.  This provides an estimate of the RP with the
highest resolution.  The flow coefficients $v_{n}^{\rb{obs}}$ are
obtained from the event averages $v_{n}^{\rb{obs}} = \langle \cos[n
(\phi - \psieone)] \rangle \, $.  The EP resolution takes the
dispersion of \psieone\ relative to \psirp\ into account, $v_{n} =
v_{n}^{\rb{obs}} / \Re_{n} \, $.  This resolution, defined as $\Re_{n}
= \langle \cos [ n (\psieone - \psirp) ] \rangle$, is determined
according to Eq.~11 in Ref.~\cite{Poskanzer:1998yz}.  Resulting values
for the resolution for flow coefficients of different order $n$ as
function of the centrality are shown in Fig.~\ref{fig:EPresolution}.

%
%

%
\begin{figure}
\includegraphics[width=1.\linewidth]{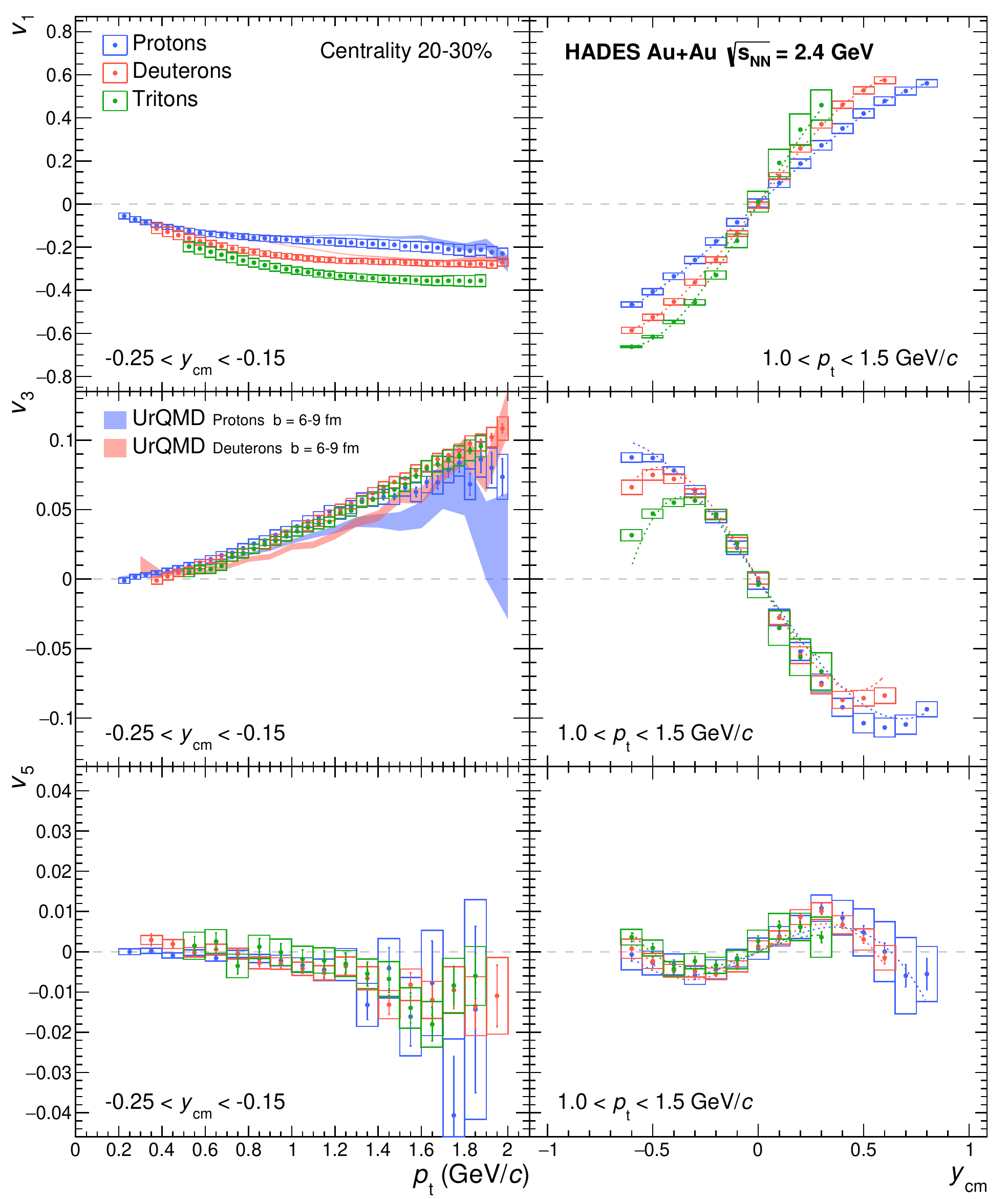}
\caption{The odd flow coefficients \vone, \vthree\ and \vfive\ for
  protons, deuterons and tritons in semi-central ($20 - 30$~\%) Au+Au
  collisions at $\sqrtsnn = 2.4$~GeV.  The left column displays the
  \pt~dependence of \vone\ (upper row), \vthree\ (middle row) and
  \vfive\ (lower row) in the rapidity interval $-0.25 < \ycm < -0.15$.
  In the right column the corresponding \ycm~dependences are
  presented.  The values are averaged over the \pt~interval $1.0 < \pt
  < 1.5$~\gevc.  The dashed coloured curves represent fits to the data
  points (see text for details).  Systematic errors are shown as open
  boxes.  UrQMD model predictions for protons and deuterons are
  depicted as shaded areas \cite{Hillmann:2019wlt}.}
\label{fig:vn_odd_cent20-30}
\end{figure}
%

%
\begin{figure}
\includegraphics[width=1.\linewidth]{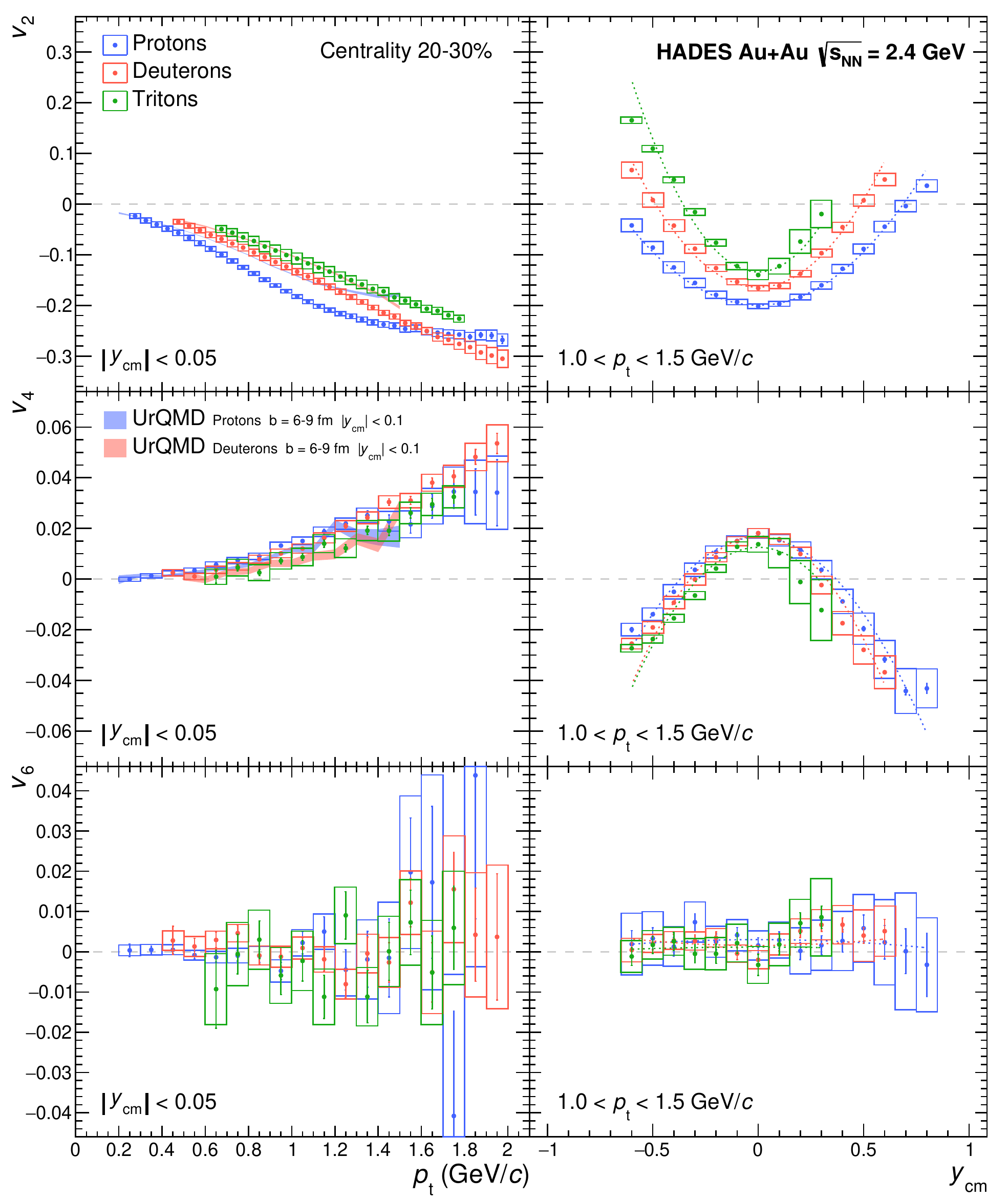}
\caption{The even flow coefficients \vtwo, \vfour\ and \vsix\ for
  protons, deuterons and tritons in semi-central ($20 - 30$~\%) Au+Au
  collisions at $\sqrtsnn = 2.4$~GeV in the same representation as in
  Fig.~\ref{fig:vn_odd_cent20-30}, except that the \pt~dependences are
  shown for the rapidity interval $|\ycm| < 0.05$.}
\label{fig:vn_even_cent20-30}
\end{figure}
%
 
%
\begin{figure}
\begin{center}
\includegraphics[width=1.\linewidth]{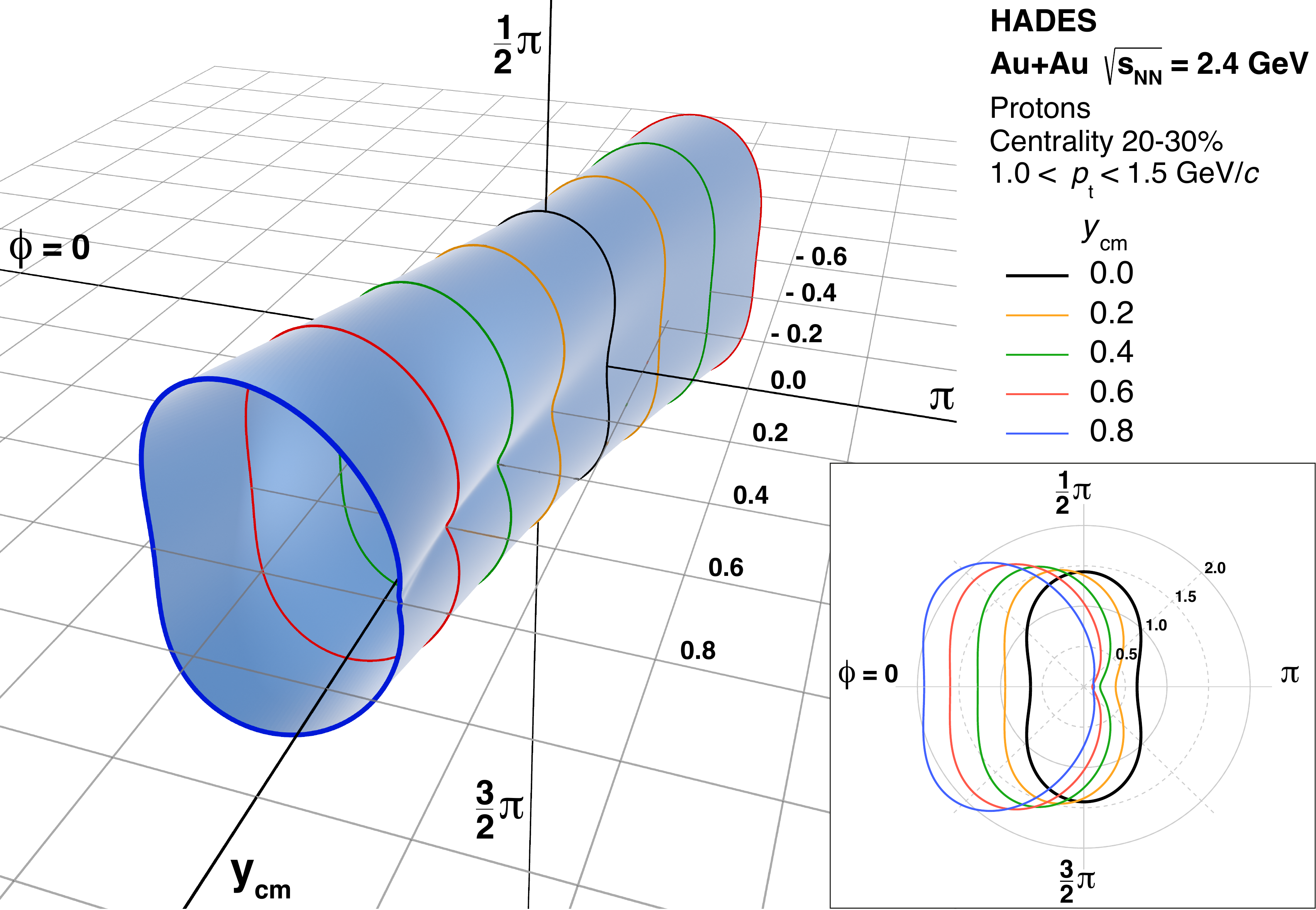}
\end{center}
\caption{A three-dimensional representation of the angular proton
  emission pattern, $1/\langle N \rangle \, (dN/d\phi)$, relative to
  the EP according to the flow coefficients of the orders $n
  = 1 - 6$, as parametrized by the fit functions shown in
  Figs.~\ref{fig:vn_odd_cent20-30} and \ref{fig:vn_even_cent20-30} for  
  semi-central ($20 - 30$~\%) Au+Au collisions.  The shape corresponds
  to the $\phi$~dependent yield normalized by the $\phi$~averaged
  value, both integrated over the \pt~interval $1.0 < \pt <
  1.5$~\gevc.  The insert presents corresponding slices at different
  forward rapidities.}
\label{fig:3DPlot}
\end{figure}
%

Systematic uncertainties of the measured flow har\-mo\-nics \vn\
result from systematic effects in the reconstruction and selection of
charged tracks, in the PID procedures, and in the corrections applied
to \vn.  They are determined separately for each particle species, the
order $n$ of the flow harmonics \vn, the centrality class and as a
function of \ycm\ and \pt\ by varying selection criteria and
parameters in the efficiency correction.  Azimuthal asymmetries due to
non-uniform acceptance and reconstruction efficiencies can cause
additional systematic uncertainties.  These are estimated by comparing
the results obtained for a fully symmetric detector (i.e. six sectors)
with those where different combinations of sectors are deliberately
excluded from the analysis.  It is found that the latter effect is
mostly dominating in case of the odd flow coefficients, while for the
even coefficients all of the above effects contribute roughly on the
same level to the point-by-point systematic uncertainties.
Furthermore, the analysis is performed on data recorded with a
reversed magnetic field setting and for each day of data taking
separately.  No significant effects are observed in these
cross-checks.  A global systematic uncertainty arises from the EP
resolution.  This is mainly caused by so-called ``non-flow''
correlations which can distort the EP measurement.  The magnitude of
these systematic effects was evaluated using the three-subevent
method, i.e. by determining the EP resolution for combinations of
different sub-events separated in rapidity, and found to be below 5~\%
for the centralities $10 - 40$~\%.

%
 
%
\begin{figure}
\includegraphics[width=1.\linewidth]{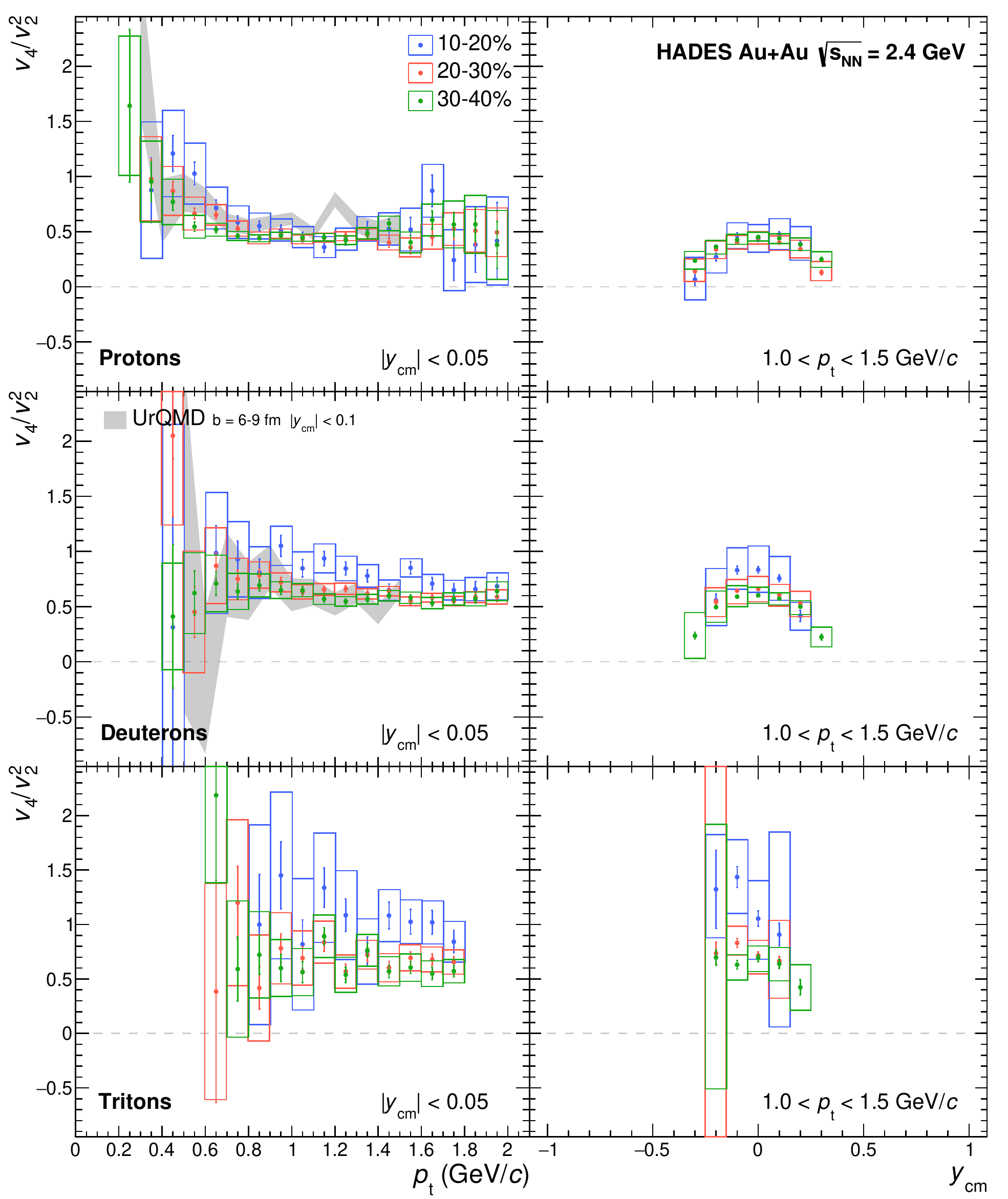}
\caption{The ratio $\vfour/\vtwo^{2}$ for protons (upper row),
  deuterons (middle row) and tritons (lower row) in Au+Au collisions
  at $\sqrtsnn = 2.4$~GeV for three different centralities.  The left
  column displays the values as a function of \pt\ at mid-rapidity
  ($|\ycm| < 0.05$) and in the right column the values averaged over
  the interval $1.0 < \pt < 1.5$~\gevc\ are shown as a function of
  rapidity.  Systematic errors are represented by open boxes.  UrQMD
  model predictions for protons and deuterons are depicted as shaded
  areas \cite{Hillmann:2019wlt}.}
\label{fig:ratio_v4v22_pdt_pt_y_allcent}
\end{figure}
%

%

Figures~\ref{fig:vn_odd_cent20-30} and \ref{fig:vn_even_cent20-30}
present an overview of the measured values for \vone\ to \vsix\ for
protons, deuterons and tritons.  Here only the values for semi-central
($20 - 30$~\%) Au+Au collisions are shown as the effect of the event
plane resolution corrections are smallest for this centrality range.
Presented is the \pt~dependence of the flow coefficients around
mid-rapidity for \vtwo, \vfour\ and \vsix, respectively at backward
rapidity for \vone, \vthree\ and \vfive, and their \ycm~dependence for
values averaged over the given \pt~interval.  The latter has been
fitted with the following functions to illustrate the symmetry of the
measurements: $v_{1,3,5}(\ycm) = a_{1,3,5}\,\ycm + b_{1,3,5} \,
\ycm^{3}$ and $v_{2,4,6}(\ycm) = c_{2,4,6} + d_{2,4,6} \, \ycm^{2}$.
The values for odd flow coefficients (\vone, \vthree\ and \vfive) are
consistent with zero at mid-rapidity, but exhibit a strong rapidity
dependence, point-symmetric around $\ycm = 0$.  Parameter \vone\
develops a prominent mass dependence ($|\vone|(\textrm{p}) <
|\vone|(\textrm{d}) < |\vone|(\textrm{t})$) when moving away from
mid-rapidity.   For larger rapidity values a mass hierarchy is also
observable for \vthree, which is, however, inverted with respect to
\vone\ ($|\vthree|(\textrm{p}) > |\vthree|(\textrm{d}) >
|\vthree|(\textrm{t})$).  In the case of \vfive, the sign of which is
opposite to the one of \vthree, no mass hierarchy can be established
due to the larger uncertainties.  For \vtwo\ around mid-rapidity a
clear mass ordering can again be observed ($|\vtwo|(\textrm{p}) >
|\vtwo|(\textrm{d}) > |\vtwo|(\textrm{t})$) up to $\pt = 1.5$~\gevc.
This mass hierarchy becomes even more pronounced when moving away from
mid-rapidity.  A similar, though less significant, mass difference is
visible for \vfour\ ($|\vfour|(\textrm{p}) > |\vfour|(\textrm{d}) >
|\vfour|(\textrm{t})$).  We note that the integrated value for \vtwo\
as measured here for protons agrees well with the world systematics,
as compiled in \cite{Andronic:2006ra,Andronic:2004cp}.  Also, we find
the same \pt~dependence of \vtwo\ at mid-rapidity as observed by FOPI
\cite{FOPI:2011aa} and KaoS \cite{Brill:1996}.  The UrQMD model is
found to provide a good description of \vone\ and \vfour\ of protons
\cite{Hillmann:2019wlt}, while discrepancies between model and data
can be observed in all other cases.

The multi-differential measurement of all flow coefficients up to
order $6$ allows to construct a three-dimensional picture of the
angular particle emission pattern relative to the RP, as
first proposed in Ref.~\cite{Voloshin:1994mz}, and is shown in
Fig.~\ref{fig:3DPlot} for the proton sample averaged over the interval
$1.0 < \pt < 1.5$~\gevc.  It is constructed by inserting values of
$v_{n}$ for a given phase space interval from the parameterizations
discussed above  (see Figs.~\ref{fig:vn_odd_cent20-30} and
\ref{fig:vn_even_cent20-30}) into the cosine of the Fourier series:
$1/\langle N \rangle \, (dN/d\phi) = 1 + 2 \sum v_{n} \cos ( n \, \phi)$.
At mid-rapidity, the combination of all flow coefficients results in a
dipole shape centered around the beam axis with the odd coefficients
being consistent with zero (see Fig.~\ref{fig:vn_odd_cent20-30}).  The
long axis of the elliptical shape is oriented along the $\phi = \pi /
2$ direction, corresponding to out-of-plane emission.  However, moving
away from mid-rapidity a more asymmetric shape appears as the
contribution of the odd coefficients increases.  As a result, at very
forward and backward rapidities the emission pattern develops a more
triangular shape.

The ratio $\vfour/\vtwo^{2}$ at mid-rapidity is shown in the left
panels of Fig.~\ref{fig:ratio_v4v22_pdt_pt_y_allcent}.  For protons a
\pt~independent value slightly below 0.5 is observed for the three
centrality intervals shown here, while for deuterons and tritons it is
found to be systematically above 0.5, both also without significant
\pt~dependence.  However, these values are only reached around
mid-rapidity as illustrated in the right panels of
Fig.~\ref{fig:ratio_v4v22_pdt_pt_y_allcent}.  A rapid drop of the
ratio is observed for the considered particle types when moving away
from mid-rapidity, as the $\ycm$~distributions of \vtwo\ and \vfour\
have different widths.  Within the semi-central range between $10$~\%
and $40$~\% no strong centrality dependence of the ratio
$\vfour/\vtwo^{2}$ is observed, as shown in
Fig.~\ref{fig:ratio_v4v22_pdt_pt_y_allcent}.  The transport model
UrQMD is found to agree well with the measured values at mid-rapidity
for protons and deuterons.  It should also be investigated whether a
description within the framework of hydrodynamic models is possible.
However, as the expected values for $\eta/s$ of dense baryonic matter
will be relatively high
\cite{Rose:2017bjz,Khvorostukhin:2010aj,Ivanov:2016hes,Demir:2008tr},
any appropriate dynamical model is expected to be far away from an
ideal fluid scenario.  As the higher order flow harmonics are here
measured relative to the first order RP, they are not
related to initial state fluctuations as is the case for higher
energies.  Thus, the geometry of the reaction system at later stages
will mainly determine the relative strength of the coefficients, which
should also be reflected in other ratios, e.g. $\vthree / (\vone
\vtwo)$.  This ratio was studied as well, however, it was found to be
dependent on \pt\ and particle type at backward rapidities, while
around mid-rapidity no reliable determination was possible.
%

%
In summary, we report a multi-differential measurement of directed,
\vone, and elliptic flow, \vtwo, and the first measurements of higher
order flow coefficients ($\vthree - \vsix$) for protons, deuterons and
tritons in heavy-ion collisions in the few GeV center-of-mass energy
regime.  All flow coefficients are determined relative to a first
order EP measured at projectile rapidities.  It is found that away
from mid-rapidity \vone\ and \vfive\ have signs opposite to the one
of \vthree, while similarly at mid-rapidity \vtwo\ is negative and
\vfour\ positive.   Combining the flow coefficients \vone~--~\vsix\
allows to construct for the first time a complete, multi-differential
picture of the emission pattern of light nuclei as a function of
rapidity and transverse momentum.  For protons at mid-rapidity the
ratio $\vfour/\vtwo^{2}$ is found to be close to a value of 0.5, while
it is slightly higher for deuterons and tritons.  A strong rapidity
dependence of this ratio is observed for all light nuclei.  Theory
calculations within a hydrodynamic framework, as e.g. described in
\cite{Russkikh:1993ct,Rischke:1995pe,Ivanov:2005yw,Karpenko:2015xea,Batyuk:2016qmb,Martinez:2019rlp},
adapted to the description of baryon dominated matter are needed to
investigate the question whether this kind of matter exhibits a
hydrodynamical behavior, at least in the last stages of the collision
prior to freeze-out.
%

%
\begin{acknowledgments}
  The collaboration gratefully acknowledges the support by
  SIP JUC Cracow, Cracow (Poland), National Science Center,
  2016/23/P/ST2/04066 POLONEZ, 2017/25/N/ST2/00580, 2017/26/M/ST2/00600;
  TU Darmstadt, Darmstadt (Germany), VH-NG-823, DFG \linebreak
  GRK 2128, DFG CRC-TR 211, BMBF:05P18RDFC1; Goethe-University,
  Frankfurt (Germany), BMBF: 06FY9100I, BMBF:05P19RFFCA, GSI F\&E, HIC
  for FAIR (LOEWE); Goethe-University, Frankfurt (Germany) and TU
  Darmstadt, Darmstadt (Germany), ExtreMe Matter Institute EMMI at GSI
  Darmstadt; TU M\"{u}nchen, Garching (Germany), MLL M\"{u}nchen,
  DFG EClust 153, GSI TMLRG1316F, BMBF 05P15WOFCA, SFB 1258, DFG
  FAB898/2-2; Russian Foundation for Basic Research (RFBR) funding
  within the research project no. 18-02-40086, National Research
  Nuclear University MEPhI in the framework of the Russian Academic
  Excellence Project (contract No.\ 02.a03.21.0005, 27.08.2013),
  Ministry of Science and Higher Education of the Russian Federation,
  Project "Fundamental properties of elementary particles and
  cosmology" No 0723-2020-0041; JLU Giessen, Giessen (Germany),
  BMBF:05P12RGGHM; IPN Orsay, Orsay Cedex (France), CNRS/IN2P3; NPI
  CAS, Rez, Rez (Czech Republic), MSMT LM2015049, OP VVV
  CZ.02.1.01/0.0/0.0/16 013/0001677, LTT17003.
\end{acknowledgments}
%

%
\bibliography{hades_flow_letter}

\begin{thebibliography}{39}%
\makeatletter
\providecommand \@ifxundefined [1]{%
 \@ifx{#1\undefined}
}%
\providecommand \@ifnum [1]{%
 \ifnum #1\expandafter \@firstoftwo
 \else \expandafter \@secondoftwo
 \fi
}%
\providecommand \@ifx [1]{%
 \ifx #1\expandafter \@firstoftwo
 \else \expandafter \@secondoftwo
 \fi
}%
\providecommand \natexlab [1]{#1}%
\providecommand \enquote  [1]{``#1''}%
\providecommand \bibnamefont  [1]{#1}%
\providecommand \bibfnamefont [1]{#1}%
\providecommand \citenamefont [1]{#1}%
\providecommand \href@noop [0]{\@secondoftwo}%
\providecommand \href [0]{\begingroup \@sanitize@url \@href}%
\providecommand \@href[1]{\@@startlink{#1}\@@href}%
\providecommand \@@href[1]{\endgroup#1\@@endlink}%
\providecommand \@sanitize@url [0]{\catcode `\\12\catcode `\$12\catcode
  `\&12\catcode `\#12\catcode `\^12\catcode `\_12\catcode `\%12\relax}%
\providecommand \@@startlink[1]{}%
\providecommand \@@endlink[0]{}%
\providecommand \url  [0]{\begingroup\@sanitize@url \@url }%
\providecommand \@url [1]{\endgroup\@href {#1}{\urlprefix }}%
\providecommand \urlprefix  [0]{URL }%
\providecommand \Eprint [0]{\href }%
\providecommand \doibase [0]{http://dx.doi.org/}%
\providecommand \selectlanguage [0]{\@gobble}%
\providecommand \bibinfo  [0]{\@secondoftwo}%
\providecommand \bibfield  [0]{\@secondoftwo}%
\providecommand \translation [1]{[#1]}%
\providecommand \BibitemOpen [0]{}%
\providecommand \bibitemStop [0]{}%
\providecommand \bibitemNoStop [0]{.\EOS\space}%
\providecommand \EOS [0]{\spacefactor3000\relax}%
\providecommand \BibitemShut  [1]{\csname bibitem#1\endcsname}%
\let\auto@bib@innerbib\@empty
\bibitem [{\citenamefont {Adamczewski-Musch}\ \emph {et~al.}(2019)\citenamefont
  {Adamczewski-Musch} \emph {et~al.}}]{Adamczewski-Musch:2019byl}%
  \BibitemOpen
  \bibfield  {author} {\bibinfo {author} {\bibfnamefont {J.}~\bibnamefont
  {Adamczewski-Musch}} \emph {et~al.} (\bibinfo {collaboration} {HADES}),\
  }\href {\doibase 10.1038/s41567-019-0583-8} {\bibfield  {journal} {\bibinfo
  {journal} {Nature Phys.}\ }\textbf {\bibinfo {volume} {15}},\ \bibinfo
  {pages} {1040} (\bibinfo {year} {2019})}\BibitemShut {NoStop}%
\bibitem [{\citenamefont {Danielewicz}\ \emph {et~al.}(2002)\citenamefont
  {Danielewicz}, \citenamefont {Lacey},\ and\ \citenamefont
  {Lynch}}]{Danielewicz:2002pu}%
  \BibitemOpen
  \bibfield  {author} {\bibinfo {author} {\bibfnamefont {P.}~\bibnamefont
  {Danielewicz}}, \bibinfo {author} {\bibfnamefont {R.}~\bibnamefont {Lacey}},
  \ and\ \bibinfo {author} {\bibfnamefont {W.~G.}\ \bibnamefont {Lynch}},\
  }\href {\doibase 10.1126/science.1078070} {\bibfield  {journal} {\bibinfo
  {journal} {Science}\ }\textbf {\bibinfo {volume} {298}},\ \bibinfo {pages}
  {1592} (\bibinfo {year} {2002})}\BibitemShut {NoStop}%
\bibitem [{\citenamefont {Le~F\`{e}vre}\ \emph {et~al.}(2016)\citenamefont
  {Le~F\`{e}vre}, \citenamefont {Leifels}, \citenamefont {Reisdorf},
  \citenamefont {Aichelin},\ and\ \citenamefont {Hartnack}}]{Fevre:2015fza}%
  \BibitemOpen
  \bibfield  {author} {\bibinfo {author} {\bibfnamefont {A.}~\bibnamefont
  {Le~F\`{e}vre}}, \bibinfo {author} {\bibfnamefont {Y.}~\bibnamefont
  {Leifels}}, \bibinfo {author} {\bibfnamefont {W.}~\bibnamefont {Reisdorf}},
  \bibinfo {author} {\bibfnamefont {J.}~\bibnamefont {Aichelin}}, \ and\
  \bibinfo {author} {\bibfnamefont {C.}~\bibnamefont {Hartnack}},\ }\href
  {\doibase 10.1016/j.nuclphysa.2015.09.015} {\bibfield  {journal} {\bibinfo
  {journal} {Nucl. Phys.}\ }\textbf {\bibinfo {volume} {A945}},\ \bibinfo
  {pages} {112} (\bibinfo {year} {2016})}\BibitemShut {NoStop}%
\bibitem [{\citenamefont {Ritter}\ and\ \citenamefont
  {Stock}(2014)}]{Ritter:2014uca}%
  \BibitemOpen
  \bibfield  {author} {\bibinfo {author} {\bibfnamefont {H.~G.}\ \bibnamefont
  {Ritter}}\ and\ \bibinfo {author} {\bibfnamefont {R.}~\bibnamefont {Stock}},\
  }\href {\doibase 10.1088/0954-3899/41/12/124002} {\bibfield  {journal}
  {\bibinfo  {journal} {J. Phys.}\ }\textbf {\bibinfo {volume} {G41}},\
  \bibinfo {pages} {124002} (\bibinfo {year} {2014})}\BibitemShut {NoStop}%
\bibitem [{\citenamefont {Andronic}\ \emph {et~al.}(2006)\citenamefont
  {Andronic}, \citenamefont {Lukasik}, \citenamefont {Reisdorf},\ and\
  \citenamefont {Trautmann}}]{Andronic:2006ra}%
  \BibitemOpen
  \bibfield  {author} {\bibinfo {author} {\bibfnamefont {A.}~\bibnamefont
  {Andronic}}, \bibinfo {author} {\bibfnamefont {J.}~\bibnamefont {Lukasik}},
  \bibinfo {author} {\bibfnamefont {W.}~\bibnamefont {Reisdorf}}, \ and\
  \bibinfo {author} {\bibfnamefont {W.}~\bibnamefont {Trautmann}},\ }\href
  {\doibase 10.1140/epja/i2006-10101-2} {\bibfield  {journal} {\bibinfo
  {journal} {Eur. Phys. J.}\ }\textbf {\bibinfo {volume} {A30}},\ \bibinfo
  {pages} {31} (\bibinfo {year} {2006})}\BibitemShut {NoStop}%
\bibitem [{\citenamefont {Herrmann}\ \emph {et~al.}(1999)\citenamefont
  {Herrmann}, \citenamefont {Wessels},\ and\ \citenamefont
  {Wienold}}]{Herrmann:1999wu}%
  \BibitemOpen
  \bibfield  {author} {\bibinfo {author} {\bibfnamefont {N.}~\bibnamefont
  {Herrmann}}, \bibinfo {author} {\bibfnamefont {J.~P.}\ \bibnamefont
  {Wessels}}, \ and\ \bibinfo {author} {\bibfnamefont {T.}~\bibnamefont
  {Wienold}},\ }\href {\doibase 10.1146/annurev.nucl.49.1.581} {\bibfield
  {journal} {\bibinfo  {journal} {Ann. Rev. Nucl. Part. Sci.}\ }\textbf
  {\bibinfo {volume} {49}},\ \bibinfo {pages} {581} (\bibinfo {year}
  {1999})}\BibitemShut {NoStop}%
\bibitem [{\citenamefont {Reisdorf}\ and\ \citenamefont
  {Ritter}(1997)}]{Reisdorf:1997fx}%
  \BibitemOpen
  \bibfield  {author} {\bibinfo {author} {\bibfnamefont {W.}~\bibnamefont
  {Reisdorf}}\ and\ \bibinfo {author} {\bibfnamefont {H.~G.}\ \bibnamefont
  {Ritter}},\ }\href {\doibase 10.1146/annurev.nucl.47.1.663} {\bibfield
  {journal} {\bibinfo  {journal} {Ann. Rev. Nucl. Part. Sci.}\ }\textbf
  {\bibinfo {volume} {47}},\ \bibinfo {pages} {663} (\bibinfo {year}
  {1997})}\BibitemShut {NoStop}%
\bibitem [{\citenamefont {Voloshin}\ and\ \citenamefont
  {Zhang}(1996)}]{Voloshin:1994mz}%
  \BibitemOpen
  \bibfield  {author} {\bibinfo {author} {\bibfnamefont {S.}~\bibnamefont
  {Voloshin}}\ and\ \bibinfo {author} {\bibfnamefont {Y.}~\bibnamefont
  {Zhang}},\ }\href {\doibase 10.1007/s002880050141} {\bibfield  {journal}
  {\bibinfo  {journal} {Z. Phys.}\ }\textbf {\bibinfo {volume} {C70}},\
  \bibinfo {pages} {665} (\bibinfo {year} {1996})}\BibitemShut {NoStop}%
\bibitem [{\citenamefont {Hillmann}\ \emph {et~al.}(2018)\citenamefont
  {Hillmann}, \citenamefont {Steinheimer},\ and\ \citenamefont
  {Bleicher}}]{Hillmann:2018nmd}%
  \BibitemOpen
  \bibfield  {author} {\bibinfo {author} {\bibfnamefont {P.}~\bibnamefont
  {Hillmann}}, \bibinfo {author} {\bibfnamefont {J.}~\bibnamefont
  {Steinheimer}}, \ and\ \bibinfo {author} {\bibfnamefont {M.}~\bibnamefont
  {Bleicher}},\ }\href {\doibase 10.1088/1361-6471/aac96f} {\bibfield
  {journal} {\bibinfo  {journal} {J. Phys.}\ }\textbf {\bibinfo {volume}
  {G45}},\ \bibinfo {pages} {085101} (\bibinfo {year} {2018})}\BibitemShut
  {NoStop}%
\bibitem [{\citenamefont {Hillmann}\ \emph {et~al.}(2020)\citenamefont
  {Hillmann}, \citenamefont {Steinheimer}, \citenamefont {Reichert},
  \citenamefont {Gaebel}, \citenamefont {Bleicher}, \citenamefont {Sombun},
  \citenamefont {Herold},\ and\ \citenamefont {Limphirat}}]{Hillmann:2019wlt}%
  \BibitemOpen
  \bibfield  {author} {\bibinfo {author} {\bibfnamefont {P.}~\bibnamefont
  {Hillmann}}, \bibinfo {author} {\bibfnamefont {J.}~\bibnamefont
  {Steinheimer}}, \bibinfo {author} {\bibfnamefont {T.}~\bibnamefont
  {Reichert}}, \bibinfo {author} {\bibfnamefont {V.}~\bibnamefont {Gaebel}},
  \bibinfo {author} {\bibfnamefont {M.}~\bibnamefont {Bleicher}}, \bibinfo
  {author} {\bibfnamefont {S.}~\bibnamefont {Sombun}}, \bibinfo {author}
  {\bibfnamefont {C.}~\bibnamefont {Herold}}, \ and\ \bibinfo {author}
  {\bibfnamefont {A.}~\bibnamefont {Limphirat}},\ }\href {\doibase
  10.1088/1361-6471/ab6fcf} {\bibfield  {journal} {\bibinfo  {journal} {J.
  Phys.}\ }\textbf {\bibinfo {volume} {G47}},\ \bibinfo {pages} {055101}
  (\bibinfo {year} {2020})}\BibitemShut {NoStop}%
\bibitem [{\citenamefont {Danielewicz}(2000)}]{Danielewicz:1999zn}%
  \BibitemOpen
  \bibfield  {author} {\bibinfo {author} {\bibfnamefont {P.}~\bibnamefont
  {Danielewicz}},\ }\href {\doibase 10.1016/S0375-9474(00)00083-X} {\bibfield
  {journal} {\bibinfo  {journal} {Nucl. Phys.}\ }\textbf {\bibinfo {volume}
  {A673}},\ \bibinfo {pages} {375} (\bibinfo {year} {2000})}\BibitemShut
  {NoStop}%
\bibitem [{\citenamefont {Heinz}\ and\ \citenamefont
  {Snellings}(2013)}]{Heinz:2013th}%
  \BibitemOpen
  \bibfield  {author} {\bibinfo {author} {\bibfnamefont {U.}~\bibnamefont
  {Heinz}}\ and\ \bibinfo {author} {\bibfnamefont {R.}~\bibnamefont
  {Snellings}},\ }\href {\doibase 10.1146/annurev-nucl-102212-170540}
  {\bibfield  {journal} {\bibinfo  {journal} {Ann. Rev. Nucl. Part. Sci.}\
  }\textbf {\bibinfo {volume} {63}},\ \bibinfo {pages} {123} (\bibinfo {year}
  {2013})}\BibitemShut {NoStop}%
\bibitem [{\citenamefont {Demir}\ and\ \citenamefont
  {Bass}(2009)}]{Demir:2008tr}%
  \BibitemOpen
  \bibfield  {author} {\bibinfo {author} {\bibfnamefont {N.}~\bibnamefont
  {Demir}}\ and\ \bibinfo {author} {\bibfnamefont {S.~A.}\ \bibnamefont
  {Bass}},\ }\href {\doibase 10.1103/PhysRevLett.102.172302} {\bibfield
  {journal} {\bibinfo  {journal} {Phys. Rev. Lett.}\ }\textbf {\bibinfo
  {volume} {102}},\ \bibinfo {pages} {172302} (\bibinfo {year}
  {2009})}\BibitemShut {NoStop}%
\bibitem [{\citenamefont {Khvorostukhin}\ \emph {et~al.}(2010)\citenamefont
  {Khvorostukhin}, \citenamefont {Toneev},\ and\ \citenamefont
  {Voskresensky}}]{Khvorostukhin:2010aj}%
  \BibitemOpen
  \bibfield  {author} {\bibinfo {author} {\bibfnamefont {A.~S.}\ \bibnamefont
  {Khvorostukhin}}, \bibinfo {author} {\bibfnamefont {V.~D.}\ \bibnamefont
  {Toneev}}, \ and\ \bibinfo {author} {\bibfnamefont {D.~N.}\ \bibnamefont
  {Voskresensky}},\ }\href {\doibase 10.1016/j.nuclphysa.2010.05.058}
  {\bibfield  {journal} {\bibinfo  {journal} {Nucl. Phys.}\ }\textbf {\bibinfo
  {volume} {A845}},\ \bibinfo {pages} {106} (\bibinfo {year}
  {2010})}\BibitemShut {NoStop}%
\bibitem [{\citenamefont {Barker}\ and\ \citenamefont
  {Danielewicz}(2019)}]{Barker:2016hqv}%
  \BibitemOpen
  \bibfield  {author} {\bibinfo {author} {\bibfnamefont {B.}~\bibnamefont
  {Barker}}\ and\ \bibinfo {author} {\bibfnamefont {P.}~\bibnamefont
  {Danielewicz}},\ }\href {\doibase 10.1103/PhysRevC.99.034607} {\bibfield
  {journal} {\bibinfo  {journal} {Phys. Rev.}\ }\textbf {\bibinfo {volume}
  {C99}},\ \bibinfo {pages} {034607} (\bibinfo {year} {2019})}\BibitemShut
  {NoStop}%
\bibitem [{\citenamefont {Rose}\ \emph {et~al.}(2018)\citenamefont {Rose},
  \citenamefont {Torres-Rincon}, \citenamefont {Sch{\"a}fer}, \citenamefont
  {Oliinychenko},\ and\ \citenamefont {Petersen}}]{Rose:2017bjz}%
  \BibitemOpen
  \bibfield  {author} {\bibinfo {author} {\bibfnamefont {J.~B.}\ \bibnamefont
  {Rose}}, \bibinfo {author} {\bibfnamefont {J.~M.}\ \bibnamefont
  {Torres-Rincon}}, \bibinfo {author} {\bibfnamefont {A.}~\bibnamefont
  {Sch{\"a}fer}}, \bibinfo {author} {\bibfnamefont {D.~R.}\ \bibnamefont
  {Oliinychenko}}, \ and\ \bibinfo {author} {\bibfnamefont {H.}~\bibnamefont
  {Petersen}},\ }\href {\doibase 10.1103/PhysRevC.97.055204} {\bibfield
  {journal} {\bibinfo  {journal} {Phys. Rev.}\ }\textbf {\bibinfo {volume}
  {C97}},\ \bibinfo {pages} {055204} (\bibinfo {year} {2018})}\BibitemShut
  {NoStop}%
\bibitem [{\citenamefont {Ivanov}\ and\ \citenamefont
  {Soldatov}(2016)}]{Ivanov:2016hes}%
  \BibitemOpen
  \bibfield  {author} {\bibinfo {author} {\bibfnamefont {{\relax Yu}.~B.}\
  \bibnamefont {Ivanov}}\ and\ \bibinfo {author} {\bibfnamefont {A.~A.}\
  \bibnamefont {Soldatov}},\ }\href {\doibase 10.1140/epja/i2016-16367-7}
  {\bibfield  {journal} {\bibinfo  {journal} {Eur. Phys. J.}\ }\textbf
  {\bibinfo {volume} {A52}},\ \bibinfo {pages} {367} (\bibinfo {year}
  {2016})}\BibitemShut {NoStop}%
\bibitem [{\citenamefont {Borghini}\ and\ \citenamefont
  {Ollitrault}(2006)}]{Borghini:2005kd}%
  \BibitemOpen
  \bibfield  {author} {\bibinfo {author} {\bibfnamefont {N.}~\bibnamefont
  {Borghini}}\ and\ \bibinfo {author} {\bibfnamefont {J.-Y.}\ \bibnamefont
  {Ollitrault}},\ }\href {\doibase 10.1016/j.physletb.2006.09.062} {\bibfield
  {journal} {\bibinfo  {journal} {Phys. Lett.}\ }\textbf {\bibinfo {volume}
  {B642}},\ \bibinfo {pages} {227} (\bibinfo {year} {2006})}\BibitemShut
  {NoStop}%
\bibitem [{\citenamefont {Adams}\ \emph {et~al.}(2004)\citenamefont {Adams}
  \emph {et~al.}}]{Adams:2003zg}%
  \BibitemOpen
  \bibfield  {author} {\bibinfo {author} {\bibfnamefont {J.}~\bibnamefont
  {Adams}} \emph {et~al.} (\bibinfo {collaboration} {STAR}),\ }\href {\doibase
  10.1103/PhysRevLett.92.062301} {\bibfield  {journal} {\bibinfo  {journal}
  {Phys. Rev. Lett.}\ }\textbf {\bibinfo {volume} {92}},\ \bibinfo {pages}
  {062301} (\bibinfo {year} {2004})}\BibitemShut {NoStop}%
\bibitem [{\citenamefont {Adare}\ \emph {et~al.}(2010)\citenamefont {Adare}
  \emph {et~al.}}]{Adare:2010ux}%
  \BibitemOpen
  \bibfield  {author} {\bibinfo {author} {\bibfnamefont {A.}~\bibnamefont
  {Adare}} \emph {et~al.} (\bibinfo {collaboration} {PHENIX}),\ }\href
  {\doibase 10.1103/PhysRevLett.105.062301} {\bibfield  {journal} {\bibinfo
  {journal} {Phys. Rev. Lett.}\ }\textbf {\bibinfo {volume} {105}},\ \bibinfo
  {pages} {062301} (\bibinfo {year} {2010})}\BibitemShut {NoStop}%
\bibitem [{\citenamefont {Aad}\ \emph {et~al.}(2012)\citenamefont {Aad} \emph
  {et~al.}}]{ATLAS:2012at}%
  \BibitemOpen
  \bibfield  {author} {\bibinfo {author} {\bibfnamefont {G.}~\bibnamefont
  {Aad}} \emph {et~al.} (\bibinfo {collaboration} {ATLAS}),\ }\href {\doibase
  10.1103/PhysRevC.86.014907} {\bibfield  {journal} {\bibinfo  {journal} {Phys.
  Rev.}\ }\textbf {\bibinfo {volume} {C86}},\ \bibinfo {pages} {014907}
  (\bibinfo {year} {2012})}\BibitemShut {NoStop}%
\bibitem [{\citenamefont {Chatrchyan}\ \emph {et~al.}(2014)\citenamefont
  {Chatrchyan} \emph {et~al.}}]{CMS:2013bza}%
  \BibitemOpen
  \bibfield  {author} {\bibinfo {author} {\bibfnamefont {S.}~\bibnamefont
  {Chatrchyan}} \emph {et~al.} (\bibinfo {collaboration} {CMS}),\ }\href
  {\doibase 10.1007/JHEP02(2014)088} {\bibfield  {journal} {\bibinfo  {journal}
  {JHEP}\ }\textbf {\bibinfo {volume} {02}},\ \bibinfo {pages} {088} (\bibinfo
  {year} {2014})}\BibitemShut {NoStop}%
\bibitem [{\citenamefont {Acharya}\ \emph {et~al.}(2018)\citenamefont {Acharya}
  \emph {et~al.}}]{Acharya:2018lmh}%
  \BibitemOpen
  \bibfield  {author} {\bibinfo {author} {\bibfnamefont {S.}~\bibnamefont
  {Acharya}} \emph {et~al.} (\bibinfo {collaboration} {ALICE}),\ }\href
  {\doibase 10.1007/JHEP07(2018)103} {\bibfield  {journal} {\bibinfo  {journal}
  {JHEP}\ }\textbf {\bibinfo {volume} {07}},\ \bibinfo {pages} {103} (\bibinfo
  {year} {2018})}\BibitemShut {NoStop}%
\bibitem [{\citenamefont {Pietraszko}\ \emph {et~al.}(2014)\citenamefont
  {Pietraszko}, \citenamefont {Galatyuk}, \citenamefont {Grilj}, \citenamefont
  {Koenig}, \citenamefont {Spataro},\ and\ \citenamefont
  {Traeger}}]{Pietraszko:2014tba}%
  \BibitemOpen
  \bibfield  {author} {\bibinfo {author} {\bibfnamefont {J.}~\bibnamefont
  {Pietraszko}}, \bibinfo {author} {\bibfnamefont {T.}~\bibnamefont
  {Galatyuk}}, \bibinfo {author} {\bibfnamefont {V.}~\bibnamefont {Grilj}},
  \bibinfo {author} {\bibfnamefont {W.}~\bibnamefont {Koenig}}, \bibinfo
  {author} {\bibfnamefont {S.}~\bibnamefont {Spataro}}, \ and\ \bibinfo
  {author} {\bibfnamefont {M.}~\bibnamefont {Traeger}},\ }\href {\doibase
  10.1016/j.nima.2014.06.006} {\bibfield  {journal} {\bibinfo  {journal} {Nucl.
  Instrum. Meth.}\ }\textbf {\bibinfo {volume} {A763}},\ \bibinfo {pages} {1}
  (\bibinfo {year} {2014})}\BibitemShut {NoStop}%
\bibitem [{\citenamefont {Adamczewski-Musch}\ \emph {et~al.}(2018)\citenamefont
  {Adamczewski-Musch} \emph {et~al.}}]{Adamczewski-Musch:2017sdk}%
  \BibitemOpen
  \bibfield  {author} {\bibinfo {author} {\bibfnamefont {J.}~\bibnamefont
  {Adamczewski-Musch}} \emph {et~al.} (\bibinfo {collaboration} {HADES}),\
  }\href {\doibase 10.1140/epja/i2018-12513-7} {\bibfield  {journal} {\bibinfo
  {journal} {Eur. Phys. J.}\ }\textbf {\bibinfo {volume} {A54}},\ \bibinfo
  {pages} {85} (\bibinfo {year} {2018})}\BibitemShut {NoStop}%
\bibitem [{\citenamefont {Agakishiev}\ \emph {et~al.}(2009)\citenamefont
  {Agakishiev} \emph {et~al.}}]{Agakishiev:2009am}%
  \BibitemOpen
  \bibfield  {author} {\bibinfo {author} {\bibfnamefont {G.}~\bibnamefont
  {Agakishiev}} \emph {et~al.} (\bibinfo {collaboration} {HADES}),\ }\href
  {\doibase 10.1140/epja/i2009-10807-5} {\bibfield  {journal} {\bibinfo
  {journal} {Eur. Phys. J.}\ }\textbf {\bibinfo {volume} {A41}},\ \bibinfo
  {pages} {243} (\bibinfo {year} {2009})}\BibitemShut {NoStop}%
\bibitem [{\citenamefont {Ollitrault}(1993)}]{Ollitrault:1993ba}%
  \BibitemOpen
  \bibfield  {author} {\bibinfo {author} {\bibfnamefont {J.-Y.}\ \bibnamefont
  {Ollitrault}},\ }\href {\doibase 10.1103/PhysRevD.48.1132} {\bibfield
  {journal} {\bibinfo  {journal} {Phys. Rev.}\ }\textbf {\bibinfo {volume}
  {D48}},\ \bibinfo {pages} {1132} (\bibinfo {year} {1993})}\BibitemShut
  {NoStop}%
\bibitem [{\citenamefont {Ollitrault}(1998)}]{Ollitrault:1997vz}%
  \BibitemOpen
  \bibfield  {author} {\bibinfo {author} {\bibfnamefont {J.-Y.}\ \bibnamefont
  {Ollitrault}},\ }\href {\doibase 10.1016/S0375-9474(98)00413-8} {\bibfield
  {journal} {\bibinfo  {journal} {Nucl. Phys.}\ }\textbf {\bibinfo {volume}
  {A638}},\ \bibinfo {pages} {195} (\bibinfo {year} {1998})}\BibitemShut
  {NoStop}%
\bibitem [{\citenamefont {Poskanzer}\ and\ \citenamefont
  {Voloshin}(1998)}]{Poskanzer:1998yz}%
  \BibitemOpen
  \bibfield  {author} {\bibinfo {author} {\bibfnamefont {A.~M.}\ \bibnamefont
  {Poskanzer}}\ and\ \bibinfo {author} {\bibfnamefont {S.~A.}\ \bibnamefont
  {Voloshin}},\ }\href {\doibase 10.1103/PhysRevC.58.1671} {\bibfield
  {journal} {\bibinfo  {journal} {Phys. Rev.}\ }\textbf {\bibinfo {volume}
  {C58}},\ \bibinfo {pages} {1671} (\bibinfo {year} {1998})}\BibitemShut
  {NoStop}%
\bibitem [{\citenamefont {Barrette}\ \emph {et~al.}(1997)\citenamefont
  {Barrette} \emph {et~al.}}]{Barrette:1997pt}%
  \BibitemOpen
  \bibfield  {author} {\bibinfo {author} {\bibfnamefont {J.}~\bibnamefont
  {Barrette}} \emph {et~al.} (\bibinfo {collaboration} {E877}),\ }\href
  {\doibase 10.1103/PhysRevC.56.3254} {\bibfield  {journal} {\bibinfo
  {journal} {Phys. Rev.}\ }\textbf {\bibinfo {volume} {C56}},\ \bibinfo {pages}
  {3254} (\bibinfo {year} {1997})}\BibitemShut {NoStop}%
\bibitem [{\citenamefont {Andronic}\ \emph {et~al.}(2005)\citenamefont
  {Andronic} \emph {et~al.}}]{Andronic:2004cp}%
  \BibitemOpen
  \bibfield  {author} {\bibinfo {author} {\bibfnamefont {A.}~\bibnamefont
  {Andronic}} \emph {et~al.} (\bibinfo {collaboration} {FOPI}),\ }\href
  {\doibase 10.1016/j.physletb.2005.02.060} {\bibfield  {journal} {\bibinfo
  {journal} {Phys. Lett.}\ }\textbf {\bibinfo {volume} {B612}},\ \bibinfo
  {pages} {173} (\bibinfo {year} {2005})}\BibitemShut {NoStop}%
\bibitem [{\citenamefont {Reisdorf}\ \emph {et~al.}(2012)\citenamefont
  {Reisdorf} \emph {et~al.}}]{FOPI:2011aa}%
  \BibitemOpen
  \bibfield  {author} {\bibinfo {author} {\bibfnamefont {W.}~\bibnamefont
  {Reisdorf}} \emph {et~al.} (\bibinfo {collaboration} {FOPI}),\ }\href
  {\doibase 10.1016/j.nuclphysa.2011.12.006} {\bibfield  {journal} {\bibinfo
  {journal} {Nucl. Phys.}\ }\textbf {\bibinfo {volume} {A876}},\ \bibinfo
  {pages} {1} (\bibinfo {year} {2012})}\BibitemShut {NoStop}%
\bibitem [{\citenamefont {Brill}\ \emph {et~al.}(1996)\citenamefont {Brill}
  \emph {et~al.}}]{Brill:1996}%
  \BibitemOpen
  \bibfield  {author} {\bibinfo {author} {\bibfnamefont {D.}~\bibnamefont
  {Brill}} \emph {et~al.} (\bibinfo {collaboration} {KaoS}),\ }\href {\doibase
  10.1007/s002180050078} {\bibfield  {journal} {\bibinfo  {journal} {Z. Phys.}\
  }\textbf {\bibinfo {volume} {A355}},\ \bibinfo {pages} {61} (\bibinfo {year}
  {1996})}\BibitemShut {NoStop}%
\bibitem [{\citenamefont {Russkikh}\ \emph {et~al.}(1994)\citenamefont
  {Russkikh}, \citenamefont {Ivanov}, \citenamefont {Pokrovsky},\ and\
  \citenamefont {Henning}}]{Russkikh:1993ct}%
  \BibitemOpen
  \bibfield  {author} {\bibinfo {author} {\bibfnamefont {V.}~\bibnamefont
  {Russkikh}}, \bibinfo {author} {\bibfnamefont {Y.}~\bibnamefont {Ivanov}},
  \bibinfo {author} {\bibfnamefont {Y.}~\bibnamefont {Pokrovsky}}, \ and\
  \bibinfo {author} {\bibfnamefont {P.}~\bibnamefont {Henning}},\ }\href
  {\doibase 10.1016/0375-9474(94)90409-X} {\bibfield  {journal} {\bibinfo
  {journal} {Nucl. Phys. A}\ }\textbf {\bibinfo {volume} {572}},\ \bibinfo
  {pages} {749} (\bibinfo {year} {1994})}\BibitemShut {NoStop}%
\bibitem [{\citenamefont {Rischke}\ \emph {et~al.}(1995)\citenamefont
  {Rischke}, \citenamefont {Pursun}, \citenamefont {Maruhn}, \citenamefont
  {Stoecker},\ and\ \citenamefont {Greiner}}]{Rischke:1995pe}%
  \BibitemOpen
  \bibfield  {author} {\bibinfo {author} {\bibfnamefont {D.~H.}\ \bibnamefont
  {Rischke}}, \bibinfo {author} {\bibfnamefont {Y.}~\bibnamefont {Pursun}},
  \bibinfo {author} {\bibfnamefont {J.~A.}\ \bibnamefont {Maruhn}}, \bibinfo
  {author} {\bibfnamefont {H.}~\bibnamefont {Stoecker}}, \ and\ \bibinfo
  {author} {\bibfnamefont {W.}~\bibnamefont {Greiner}},\ }\href@noop {}
  {\bibfield  {journal} {\bibinfo  {journal} {Acta Phys. Hung. A}\ }\textbf
  {\bibinfo {volume} {1}},\ \bibinfo {pages} {309} (\bibinfo {year}
  {1995})}\BibitemShut {NoStop}%
\bibitem [{\citenamefont {Ivanov}\ \emph {et~al.}(2006)\citenamefont {Ivanov},
  \citenamefont {Russkikh},\ and\ \citenamefont {Toneev}}]{Ivanov:2005yw}%
  \BibitemOpen
  \bibfield  {author} {\bibinfo {author} {\bibfnamefont {{\relax Yu}.~B.}\
  \bibnamefont {Ivanov}}, \bibinfo {author} {\bibfnamefont {V.~N.}\
  \bibnamefont {Russkikh}}, \ and\ \bibinfo {author} {\bibfnamefont {V.~D.}\
  \bibnamefont {Toneev}},\ }\href {\doibase 10.1103/PhysRevC.73.044904}
  {\bibfield  {journal} {\bibinfo  {journal} {Phys. Rev.}\ }\textbf {\bibinfo
  {volume} {C73}},\ \bibinfo {pages} {044904} (\bibinfo {year}
  {2006})}\BibitemShut {NoStop}%
\bibitem [{\citenamefont {Karpenko}\ \emph {et~al.}(2015)\citenamefont
  {Karpenko}, \citenamefont {Huovinen}, \citenamefont {Petersen},\ and\
  \citenamefont {Bleicher}}]{Karpenko:2015xea}%
  \BibitemOpen
  \bibfield  {author} {\bibinfo {author} {\bibfnamefont {I.}~\bibnamefont
  {Karpenko}}, \bibinfo {author} {\bibfnamefont {P.}~\bibnamefont {Huovinen}},
  \bibinfo {author} {\bibfnamefont {H.}~\bibnamefont {Petersen}}, \ and\
  \bibinfo {author} {\bibfnamefont {M.}~\bibnamefont {Bleicher}},\ }\href
  {\doibase 10.1103/PhysRevC.91.064901} {\bibfield  {journal} {\bibinfo
  {journal} {Phys. Rev. C}\ }\textbf {\bibinfo {volume} {91}},\ \bibinfo
  {pages} {064901} (\bibinfo {year} {2015})}\BibitemShut {NoStop}%
\bibitem [{\citenamefont {Batyuk}\ \emph {et~al.}(2016)\citenamefont {Batyuk},
  \citenamefont {Blaschke}, \citenamefont {Bleicher}, \citenamefont {Ivanov},
  \citenamefont {Karpenko}, \citenamefont {Merts}, \citenamefont {Nahrgang},
  \citenamefont {Petersen},\ and\ \citenamefont
  {Rogachevsky}}]{Batyuk:2016qmb}%
  \BibitemOpen
  \bibfield  {author} {\bibinfo {author} {\bibfnamefont {P.}~\bibnamefont
  {Batyuk}}, \bibinfo {author} {\bibfnamefont {D.}~\bibnamefont {Blaschke}},
  \bibinfo {author} {\bibfnamefont {M.}~\bibnamefont {Bleicher}}, \bibinfo
  {author} {\bibfnamefont {{\relax Yu}.~B.}\ \bibnamefont {Ivanov}}, \bibinfo
  {author} {\bibfnamefont {I.}~\bibnamefont {Karpenko}}, \bibinfo {author}
  {\bibfnamefont {S.}~\bibnamefont {Merts}}, \bibinfo {author} {\bibfnamefont
  {M.}~\bibnamefont {Nahrgang}}, \bibinfo {author} {\bibfnamefont
  {H.}~\bibnamefont {Petersen}}, \ and\ \bibinfo {author} {\bibfnamefont
  {O.}~\bibnamefont {Rogachevsky}},\ }\href {\doibase
  10.1103/PhysRevC.94.044917} {\bibfield  {journal} {\bibinfo  {journal} {Phys.
  Rev.}\ }\textbf {\bibinfo {volume} {C94}},\ \bibinfo {pages} {044917}
  (\bibinfo {year} {2016})}\BibitemShut {NoStop}%
\bibitem [{\citenamefont {Martinez}\ \emph {et~al.}(2019)\citenamefont
  {Martinez}, \citenamefont {Sievert}, \citenamefont {Wertepny},\ and\
  \citenamefont {Noronha-Hostler}}]{Martinez:2019rlp}%
  \BibitemOpen
  \bibfield  {author} {\bibinfo {author} {\bibfnamefont {M.}~\bibnamefont
  {Martinez}}, \bibinfo {author} {\bibfnamefont {M.~D.}\ \bibnamefont
  {Sievert}}, \bibinfo {author} {\bibfnamefont {D.~E.}\ \bibnamefont
  {Wertepny}}, \ and\ \bibinfo {author} {\bibfnamefont {J.}~\bibnamefont
  {Noronha-Hostler}},\ }\href@noop {} {\  (\bibinfo {year} {2019})},\ \Eprint
  {http://arxiv.org/abs/1911.12454} {arXiv:1911.12454 [nucl-th]} \BibitemShut
  {NoStop}%
\end{thebibliography}%
%

\end{document}